\documentclass[conference]{IEEEtran}
\IEEEoverridecommandlockouts
\usepackage{cite}
\usepackage{amsmath,amssymb,amsfonts}
\usepackage{algorithmic}
\usepackage{adjustbox}
\usepackage{lscape}
\usepackage{textcomp}
\usepackage{xcolor}
\usepackage{hyperref}
\usepackage{float}
\usepackage{listings} 

\def\BibTeX{{\rm B\kern-.05em{\sc i\kern-.025em b}\kern-.08em
    T\kern-.1667em\lower.7ex\hbox{E}\kern-.125emX}}
\usepackage{graphicx} 
\begin{document}

\title{ Secure coding for web applications: Frameworks, challenges, and the role of LLMs
}

    \author{
    Kiana Kiashemshaki\textsuperscript{1}, Mohammad Jalili Torkamani\textsuperscript{2}, Negin Mahmoudi\textsuperscript{3}\\
    \textsuperscript{1}Department of Computer Science, Bowling Green State University, Bowling Green, OH, USA \\
    \textsuperscript{2}School of Computing,  University of Nebraska-Lincoln,  Lincoln, Nebraska, USA \\
    \textsuperscript{3} Department of Civil Environmental and Ocean Engineering, Stevens Institute of Technology, New Jersey , USA \\
       Emails: kkiana@bgsu.edu,
       mJaliliTorkamani2@huskers.unl.edu,
       nmahmoud1@stevens.edu
}

\maketitle

\begin{abstract}
Secure coding is a critical yet often overlooked practice in software development. Despite extensive awareness efforts, real-world adoption remains inconsistent due to organizational, educational, and technical barriers. This paper provides a comprehensive review of secure coding practices across major frameworks and domains, including web development, DevSecOps, and cloud security. It introduces a structured framework comparison and categorizes threats aligned with the OWASP Top 10. Additionally, we explore the rising role of Large Language Models (LLMs) in evaluating and recommending secure code, presenting a reproducible case study across four major vulnerability types. This paper offers practical insights for researchers, developers, and educators on integrating secure coding into real-world development processes.
\end{abstract}


\section{Introduction}

\IEEEPARstart{T}he widespread adoption of web applications has dramatically reshaped digital service delivery across industries. From finance and e-commerce to healthcare and education, organizations increasingly rely on web based systems to streamline operations, deliver services, and engage with users. However, this growing dependency has made web applications one of the most targeted vectors for cyberattacks, exposing sensitive data and critical infrastructure to significant risk. As a result, embedding robust security mechanisms directly into the software development lifecycle has become an essential component of modern software engineering~\cite{b2,b6}.

In recent years, the frequency, scale, and sophistication of web based attacks have surged. Common vulnerabilities such as SQL injection (SQLi), cross-site scripting (XSS), broken authentication, and cross-site request forgery (CSRF) continue to dominate global threat landscapes~\cite{b9,b12,b14}. These flaws are often the result of insecure coding practices, weak input validation, and a lack of secure software design principles. Empirical studies have shown that a large percentage of data breaches can be traced back to such preventable vulnerabilities~\cite{b14}.

To mitigate these risks, several secure coding frameworks and best practice guidelines have been proposed. The OWASP Top 10 identifies the most critical security risks for web applications and provides actionable strategies to address them~\cite{b6}. The NIST Cybersecurity Framework offers a structured, high level approach to managing cybersecurity risk through its five core functions: identify, protect, detect, respond, and recover~\cite{b7}. Similarly, the Secure Software Development Lifecycle (SSDLC) emphasizes the integration of security controls at every stage of the development process, from requirements gathering to deployment and maintenance~\cite{b2,b4}.

Despite the availability of such resources, the consistent implementation of secure coding practices remains a challenge in the real world. Many organizations face barriers such as inadequate developer training, pressure to meet tight deadlines, and limited security resources. This often results in a prioritization of feature delivery over security assurance, leaving applications vulnerable to exploitation~\cite{b1,b3,b5}.

Furthermore, the rapid evolution of software development paradigms such as agile methodologies, DevOps pipelines, and cloud native environments has introduced new complexities that traditional security practices struggle to keep up with. These shifts necessitate a more adaptive, automated, and integrated approach to secure coding, aligning security measures with modern development workflows.

This paper presents a comprehensive review of secure coding frameworks and practices in the context of web application security. It explores established standards, analyzes key implementation challenges, and investigates emerging trends such as AI powered security tools, zero trust architectures, and security as code in DevSecOps environments. By synthesizing insights from academic literature, industry reports, and real-world practices, the paper aims to offer practical guidance for developers, security engineers, and organizational leaders striving to enhance software security.

\begin{figure}[ht]
    \centering
    \includegraphics[width=0.9\linewidth]{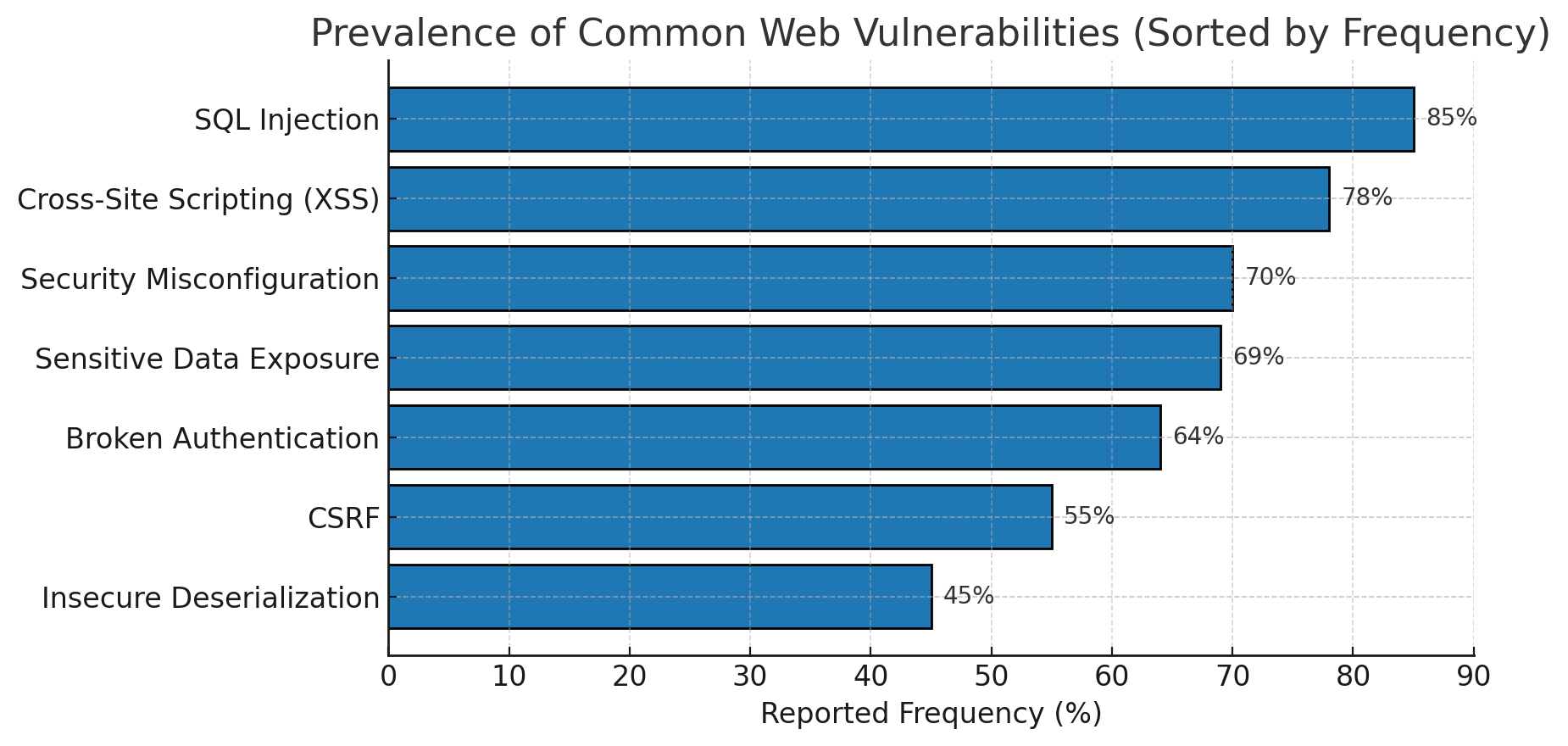}
    \caption{Prevalence of Common Web Vulnerabilities (adapted from OWASP Top 10 and industry surveys).}
    \label{fig:web_vuln}
\end{figure}

As illustrated in Figure~\ref{fig:web_vuln}, injection flaws and XSS vulnerabilities continue to be among the most commonly reported security issues in modern web applications.

\section{LITERATURE REVIEW}

Secure coding has long been recognized as a critical discipline in the field of cybersecurity. Numerous studies have examined the causes of software vulnerabilities and proposed various frameworks, strategies, and methodologies to prevent them. This section presents a structured overview of the key literature, organized around core contributions: proactive security integration, vulnerability identification and mitigation, automation, and emerging security technologies.

\subsection{Proactive Integration of Security in the Development Lifecycle}

Early efforts in secure software design emphasized the importance of integrating security from the initial stages of development. Shostack~\cite{b17} introduced threat modeling as a proactive methodology, urging developers to assess risks during the design phase rather than waiting until deployment. His approach underscored that identifying and mitigating threats early reduces the cost and complexity of security interventions.

Similarly, McGraw~\cite{b2} advocated for building security into the software lifecycle by adopting risk based security assessments, secure architecture principles, and developer education. He positioned security not as an add on but as a foundational element in software design. Lipner~\cite{b18} demonstrated the practical implementation of these ideas through Microsoft’s Security Development Lifecycle (SDL), showing that introducing security checkpoints throughout development can reduce vulnerabilities and enhance software reliability.

\subsection{Vulnerability Patterns and Defensive Coding Practices}

A Complementary body of research focused on understanding how specific coding flaws lead to exploitation. Howard and LeBlanc~\cite{b1} analyzed real-world case studies of security failures and proposed coding strategies to avoid common mistakes. They emphasized consistency in applying secure coding principles across teams and projects.

Viega and McGraw~\cite{b4} further highlighted typical software vulnerabilities such as poor input validation, weak error handling, and insecure authentication. They promoted defensive programming, encouraging developers to assume that any input or dependency might be malicious and to write code accordingly.

Whittaker and Thompson~\cite{b19} contributed by mapping attack patterns to software defects. Their work offered insight into how attackers exploit insecure code and how developers can anticipate such attacks during development. This mindset shift from reactive defense to attacker aware development laid the groundwork for more resilient applications.

\subsection{Security Testing and Automation}

Another significant area of research examines tools and techniques for identifying security flaws during development and testing. Jones and Rastogi~\cite{b20} compared static and dynamic analysis tools, showing that both approaches have strengths and limitations. Static analysis is valuable for early code level checks, while dynamic analysis is critical for detecting runtime vulnerabilities. The authors recommended a hybrid approach to maximize security coverage.

Sharma et al.~\cite{b21} expanded on this by exploring AI powered vulnerability scanners that automate threat detection. Their study found that machine learning based tools improve scanning speed and efficiency, especially in large codebases. However, they also noted the risk of false positives, reinforcing the need for expert validation.

\subsection{Evaluating Frameworks and Emerging Technologies}

Rajput et al.~\cite{b22} evaluated several secure coding frameworks, including OWASP and NIST, across diverse software development environments. Their findings revealed that while these frameworks offer robust guidelines, adoption is often hindered by limited expertise, resource constraints, and organizational resistance.

To address security challenges in decentralized and distributed systems, Ghobadi and Tavana~\cite{b23} proposed blockchain based authentication mechanisms. They argued that decentralized identity management can reduce the risks of credential theft and unauthorized access in web applications.

\subsection{Summary of Literature Insights}

Taken together, the literature shows that secure coding is a multi dimensional field that evolves in response to both technical and organizational factors. Effective approaches combine early integration of security practices~\cite{b2,b17}, use of structured frameworks~\cite{b6,b7,b22}, automation tools~\cite{b21}, and support for emerging technologies such as blockchain~\cite{b23}. While there is no one size fits all solution, the convergence of best practices from software engineering, cybersecurity, and AI research continues to shape the future of secure development.

\begin{figure}[hbt!]
    \centering
    \includegraphics[width=\linewidth]{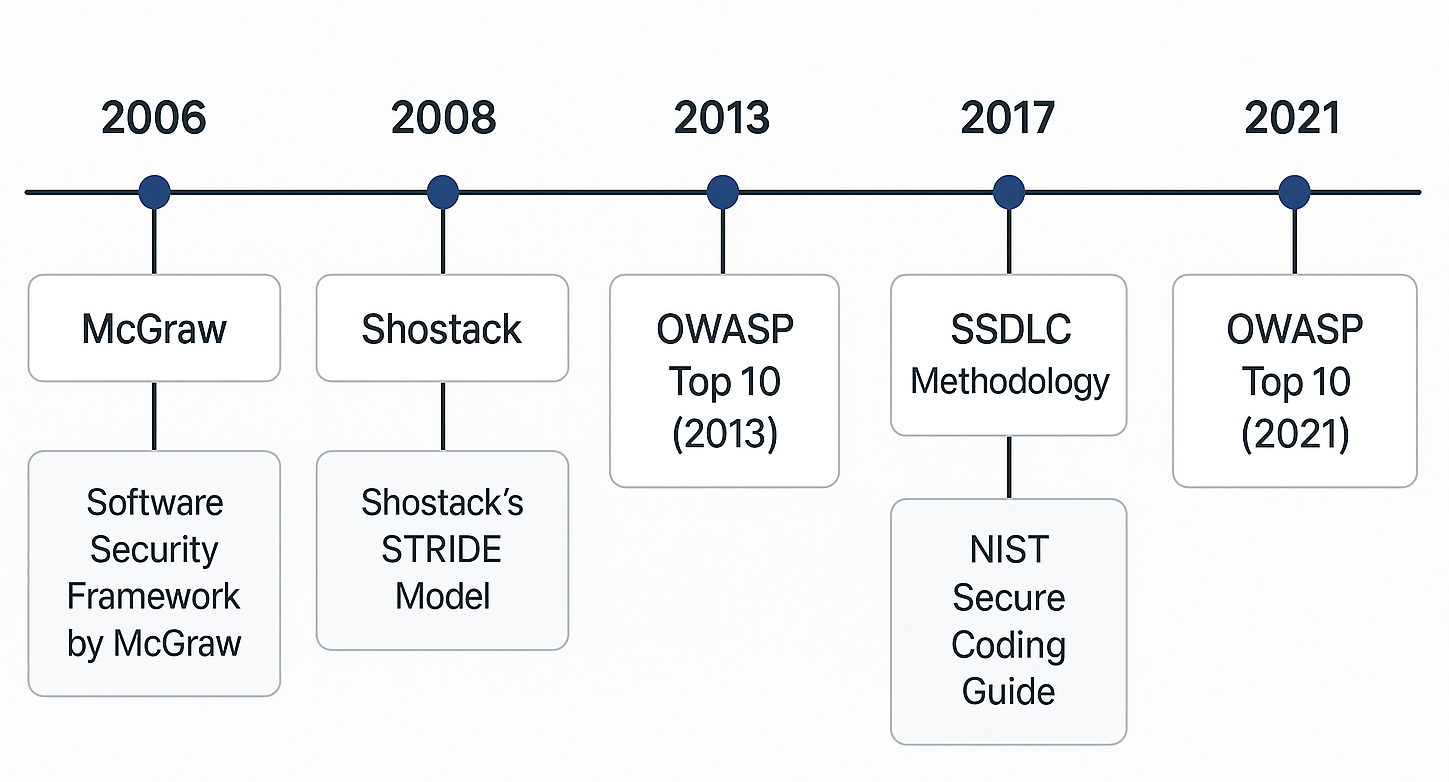}
    \caption{Timeline of Secure Coding Frameworks and Milestones.}
    \label{fig:secure_coding_timeline}
\end{figure}

To illustrate how secure coding concepts and frameworks have evolved over time, Figure~\ref{fig:secure_coding_timeline} presents a timeline of key milestones. These include foundational works such as McGraw’s secure development lifecycle, Shostack’s introduction of threat modeling, and the iterative updates to the OWASP Top 10, as well as the formalization of NIST’s cybersecurity framework and more recent AI driven developments.

Secure coding remains central to preventing critical threats like SQL injection, XSS, and broken authentication~\cite{b2,b4,b18}. However, the literature also makes it clear that technical strategies alone are insufficient developer training, organizational buy in, and continuous adaptation to new risks are equally vital for lasting impact.

\section{REMARKS ON LITERATURE REVIEW}

The literature on secure coding offers a wealth of insights into foundational practices, tools, and frameworks. However, several key challenges and research gaps emerge that require further investigation to bridge the divide between theory and practice. This section highlights three major observations derived from the reviewed works: challenges in implementation, gaps in research coverage, and limitations of automated security testing.

\subsection*{A. Challenges in the Practical Adoption of Secure Coding}

Despite the presence of structured guidelines and mature frameworks, the consistent and effective implementation of secure coding practices remains elusive in many real-world settings. Rajput et al.~\cite{b22} identified several obstacles that hinder adoption, including insufficient developer training, lack of management commitment, and competing business priorities. These organizational factors often result in security being deprioritized in favor of faster delivery timelines.

Whittaker and Thompson~\cite{b19} further emphasized that many security failures are not purely technical but arise from developers' limited understanding of attacker behaviors and threat vectors. Without adequate training and attacker centric thinking, even well intentioned teams may overlook exploitable flaws, leading to persistent vulnerabilities in deployed software.

\subsection*{B. Research Gaps in Secure Coding Education and Integration}

While numerous studies propose technical solutions for improving code security, far fewer address the systemic barriers to their adoption particularly in educational and organizational contexts. Ghobadi and Tavana~\cite{b23} observed that secure coding principles are not consistently embedded in software engineering curricula, leaving new graduates unprepared to handle security responsibilities in practice.

Moreover, Shostack~\cite{b17} highlighted the lack of alignment between threat modeling approaches and modern agile or DevOps workflows. These fast paced development environments often omit early stage security planning due to time constraints and tooling complexity, which limits the effectiveness of traditional models like SDL and SSDLC.

\subsection*{C. Limitations of Automated Testing and AI based Security Tools}

With the increasing complexity of software systems, security testing has evolved to incorporate automation and artificial intelligence. Sharma et al.~\cite{b21} demonstrated that AI powered static and dynamic analysis tools can identify common vulnerabilities more quickly and comprehensively than manual methods, making them valuable components of modern development pipelines.

However, these tools are not without limitations. Automated scanners frequently generate false positives, lack contextual understanding, and may struggle with complex, logic based vulnerabilities. As a result, expert oversight is still essential to validate findings and guide remediation efforts. Over reliance on automation without human review can lead to misplaced confidence and overlooked risks.

\subsection*{D. Summary of Literature Remarks}

Taken together, these observations highlight a clear disconnect between secure coding theory and its operationalization in practice. Technical tools and frameworks exist, but their real-world impact is often constrained by educational, cultural, and organizational factors. The literature suggests that future research should not only focus on advancing technical solutions but also prioritize improved training, better integration with agile methodologies, and hybrid approaches that balance automation with human expertise.

\section{COMPARISON OF SECURE CODING FRAMEWORKS}

Several well established frameworks have emerged to guide secure coding efforts and improve the security posture of software systems. This section compares three of the most influential models: the OWASP Top 10, the NIST Cybersecurity Framework (CSF), and the Secure Software Development Lifecycle (SSDLC). Each serves a distinct role in promoting secure development, with unique strengths and limitations.

\subsection*{A. OWASP Top 10 and Secure Coding}

The OWASP Top 10 is a widely recognized industry benchmark that outlines the most critical security risks to web applications~\cite{b6}. It covers issues such as injection attacks, broken access control, and misconfigurations. The list is updated regularly most recently in 2021 to reflect emerging threats and trends, and it is mapped to Common Weakness Enumerations (CWEs), providing actionable guidance for developers.

Howard and LeBlanc emphasize that the OWASP Top 10 is particularly effective in encouraging secure input handling and validation, which helps mitigate frequent attack vectors~\cite{b1}. Similarly, McGraw supports OWASP’s emphasis on incorporating security earlier in the development process, reinforcing its role as a proactive risk reduction strategy~\cite{b2}.

One key strength of the OWASP Top 10 lies in its simplicity and accessibility it translates complex security concerns into ten high impact categories that are easily understandable even for non security professionals. However, its limitations are equally noteworthy. The Top 10 is focused exclusively on web application risks and lacks depth in areas such as infrastructure security or CMS specific threats~\cite{b29}. Additionally, its high level nature may leave organizations without sufficient guidance for implementation.

\subsection*{B. NIST Cybersecurity Framework}

The NIST CSF offers a broad, adaptable structure for managing cybersecurity risk across organizational and technical domains~\cite{b7}. It defines six core functions govern, identify, protect, detect, respond, and recover that serve as a comprehensive roadmap for integrating security into both strategic planning and operational workflows. Seacord highlights that adopting this framework improves alignment between development and security teams while promoting robust software engineering practices~\cite{b3}.

The flexibility of the NIST framework is one of its greatest strengths. It can be customized to fit an organization's size, industry, and maturity level, making it suitable for both small businesses and large enterprises~\cite{b30}. Moreover, its adoption supports regulatory compliance and improves stakeholder confidence.

However, the framework’s high level of abstraction may pose challenges. It does not prescribe specific technical controls, which can hinder implementation in organizations lacking in house cybersecurity expertise~\cite{b31}. This often leads to inconsistent adoption unless supported by strong governance and training programs.

\begin{figure}[H]
    \centering
    \includegraphics[width=0.48\textwidth]{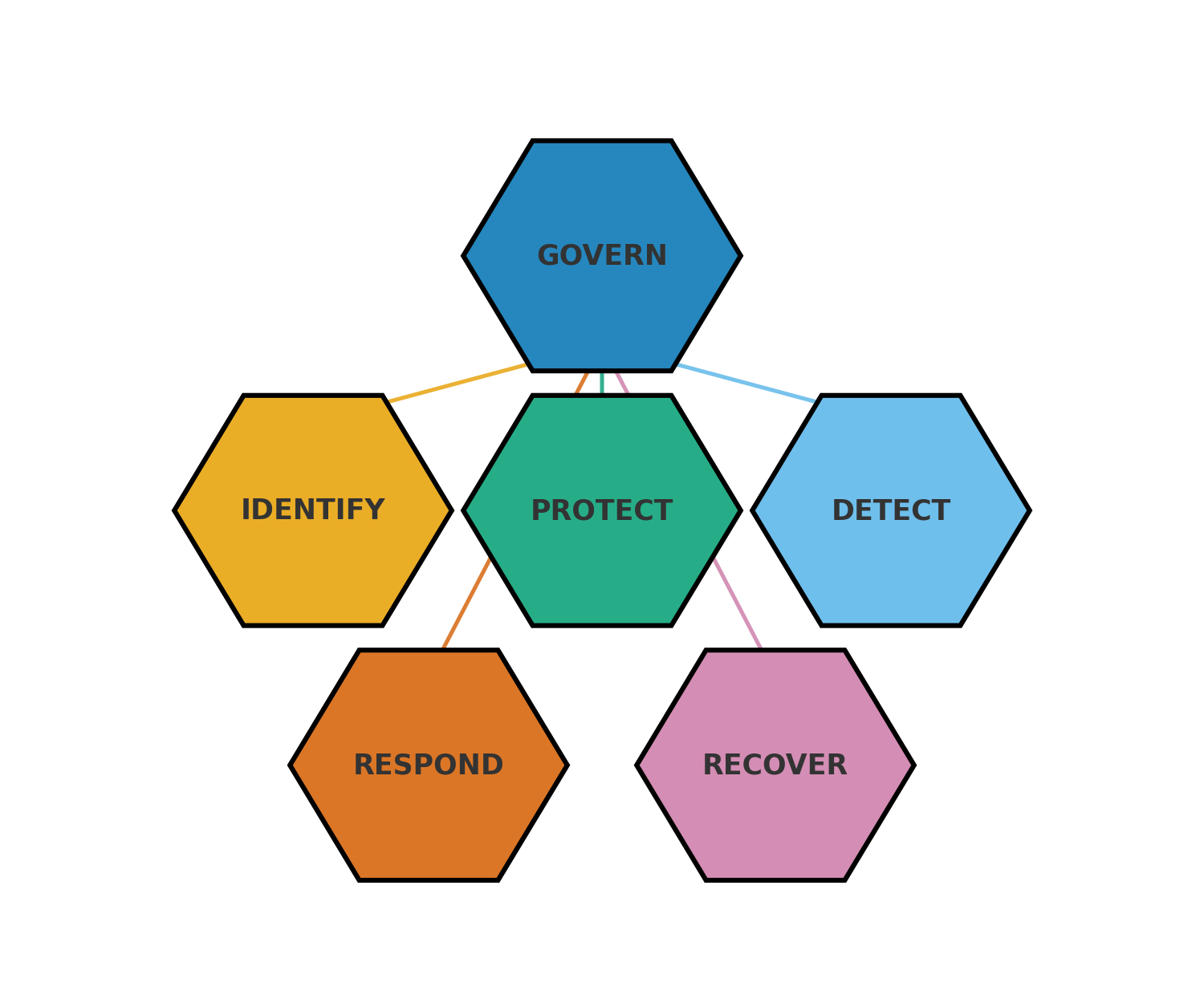}
    \caption{NIST Cybersecurity Framework.}
    \label{fig:nist_framework}
\end{figure}

Figure~\ref{fig:nist_framework} provides a visual overview of the NIST CSF's core functions.

\subsection*{C. Secure Software Development Lifecycle (SSDLC)}

The SSDLC emphasizes integrating security activities throughout all phases of software development from requirements gathering and design to implementation, testing, and maintenance. By embedding security into each phase, the SSDLC model promotes early identification and mitigation of vulnerabilities, leading to more secure and reliable software~\cite{b4}.

Viega and McGraw~\cite{b4} and Long et al.~\cite{b5} stress that this early integration significantly reduces downstream security costs and improves code quality. The SSDLC is particularly beneficial in reducing overlooked edge cases, ensuring more thorough validation and testing, and building a security conscious development culture~\cite{b32}.

Despite its strengths, SSDLC introduces additional complexity and demands greater coordination across teams. In agile and fast paced development environments, integrating security in every iteration can be seen as time consuming, which sometimes results in resistance from developers or project managers.

\begin{figure}[H]
    \centering
    \includegraphics[width=0.95\linewidth]{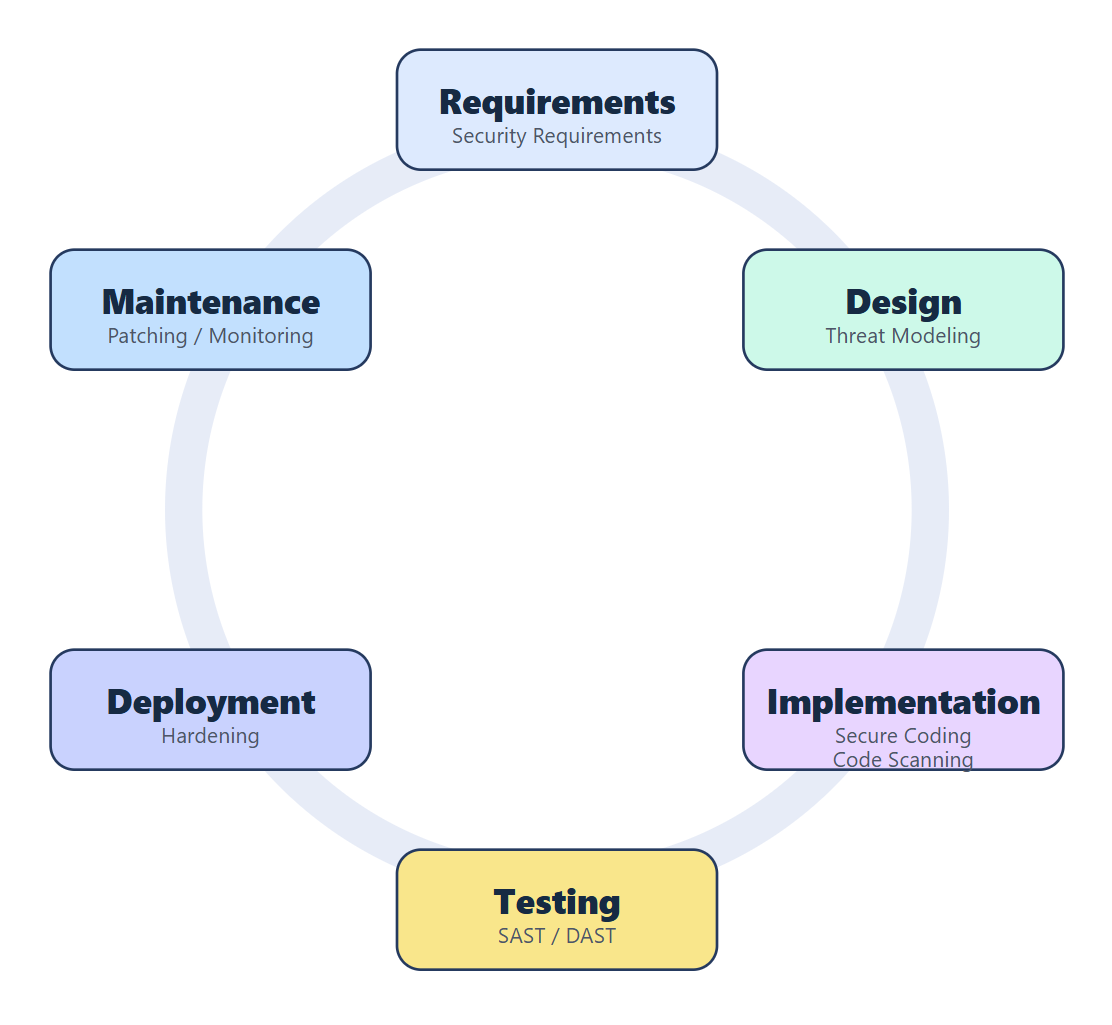}
    \caption{Secure Software Development Life Cycle (SSDLC).}
    \label{fig:ssdlc}
\end{figure}

Figure~\ref{fig:ssdlc} illustrates how security tasks are embedded within each stage of the SSDLC process.

\subsection*{D. Comparative Analysis}

Each of the three frameworks plays a vital role in enhancing software security, but they differ in scope, focus, and implementation difficulty. Table~\ref{tab:frameworks} summarizes their key characteristics, offering a side by side comparison of their purposes, advantages, and practical limitations.

\begin{table*}[hbt!]
\centering
\caption{Comparison of Secure Coding Frameworks.}
\label{tab:frameworks}
\resizebox{\textwidth}{!}{%
\begin{tabular}{|l|p{13cm}|}
\hline
\textbf{Framework} & \textbf{Purpose, Strengths, and Limitations} \\ \hline
\textbf{OWASP Top 10} & 
\textbf{Purpose:} Identify and prioritize top web application vulnerabilities. \newline
\textbf{Strengths:} Easy to understand, regularly updated, widely adopted in training. \newline
\textbf{Limitations:} High level; limited guidance for implementation and infrastructure threats. \\ \hline
\textbf{NIST CSF} & 
\textbf{Purpose:} Provide a flexible framework for managing cybersecurity risk. \newline
\textbf{Strengths:} Broad applicability; supports governance, compliance, and scalability. \newline
\textbf{Limitations:} Too abstract for direct use by developers; requires security expertise. \\ \hline
\textbf{SSDLC} & 
\textbf{Purpose:} Integrate security into each stage of software development. \newline
\textbf{Strengths:} Proactive, improves code robustness, identifies flaws early. \newline
\textbf{Limitations:} Resource intensive; may slow development and be resisted in agile environments. \\ \hline
\end{tabular}%
}
\end{table*}

\section{CHALLENGES IN IMPLEMENTING SECURE CODING PRACTICES}

Although secure coding frameworks and tools are increasingly available, organizations still face numerous barriers to consistent adoption. These challenges span organizational constraints, educational gaps, and technical limitations. This section categorizes the core difficulties into three key dimensions.

\subsection*{A. Organizational Challenges}

\subsubsection*{1) Resource and Budget Constraints}
Small to mid sized enterprises often operate under limited budgets and with minimal personnel, making it difficult to prioritize secure coding practices. These constraints hinder investment in advanced security tools, code review processes, or developer training~\cite{b9}. Howard and LeBlanc~\cite{b1} note that organizations frequently underestimate the long term costs of insecure development, failing to allocate adequate resources for prevention strategies.

\subsubsection*{2) Limited Management Support}
The successful implementation of secure development policies often requires buy in from senior leadership. However, decision makers frequently prioritize speed, feature delivery, and cost efficiency over long term security~\cite{b24}. This strategic misalignment causes security to be treated as a secondary concern, leading developers to deprioritize security in their day to day workflows.

\subsubsection*{3) “Speed to Market” Culture}
In highly competitive markets, the pressure to rapidly release software often supersedes the focus on building secure systems~\cite{b17}. This urgency fosters a reactive rather than proactive approach to security, where controls are added post deployment if at all. Such practices increase the likelihood of releasing software with untested vulnerabilities~\cite{b6,b17}.

\subsection*{B. Educational and Knowledge Gaps}

\subsubsection*{1) Inadequate Formal Training}
A major barrier to secure software development is the lack of formal education in security principles among developers. Many computer science and software engineering programs continue to overlook secure coding as a core component of the curriculum~\cite{b10,b3}. As McGraw argues, developers without foundational knowledge in secure development may inadvertently introduce exploitable flaws~\cite{b2}.

\subsubsection*{2) Lack of Developer Awareness}
Even when resources are available, many developers remain unaware of best practices or credible training programs. The Open Source Security Foundation (OpenSSF) reports that 53\% of developers have never received training in secure coding practices~\cite{b25}. This is especially common among self taught developers, who may not have been exposed to security standards in a structured way.

\subsubsection*{3) Limited On the Job Training}
Beyond initial education, there is a lack of continuous professional development in application security. Many organizations do not offer routine security training or allocate time for developers to stay informed about evolving threats. According to recent surveys, over half of developers attribute their inability to build secure applications to insufficient in house training programs.

\subsection*{C. Technical Challenges}

\subsubsection*{1) Legacy Systems and Technical Debt}
Large organizations often maintain legacy systems with outdated architectures and codebases that were not designed with modern security in mind~\cite{b26}. Modifying such systems to align with current best practices is resource intensive and risky, especially when existing functionality is poorly documented. This technical debt discourages developers from making necessary security updates, leaving the systems exposed.

\subsubsection*{2) Inadequate Testing and Scanning}
Comprehensive security testing is frequently overlooked under time constraints. Dynamic analysis, penetration testing, and manual code reviews are time consuming and are often omitted to meet delivery deadlines. In a 2024 survey by Contrast Security, 53\% of organizations admitted to skipping security scans to expedite releases~\cite{b27}.

\subsubsection*{3) Architectural Complexity}
Modern applications are composed of interconnected microservices, third party APIs, cloud native resources, and open source dependencies. This heterogeneous architecture significantly increases the attack surface and makes it difficult to enforce consistent security practices across all components~\cite{b26}. Managing security across such diverse and distributed environments requires advanced tooling, centralized policies, and cross team coordination all of which are often lacking.

\subsection*{D. Summary of Challenges}

In summary, the challenges in implementing secure coding are multifaceted. Organizational inertia, gaps in developer knowledge, and the inherent complexity of modern software systems all contribute to inconsistent adoption of best practices. Addressing these barriers requires a holistic strategy that combines education, tooling, process redesign, and executive level commitment.

\begin{figure}[H]
    \centering
    \includegraphics[width=0.5\textwidth]{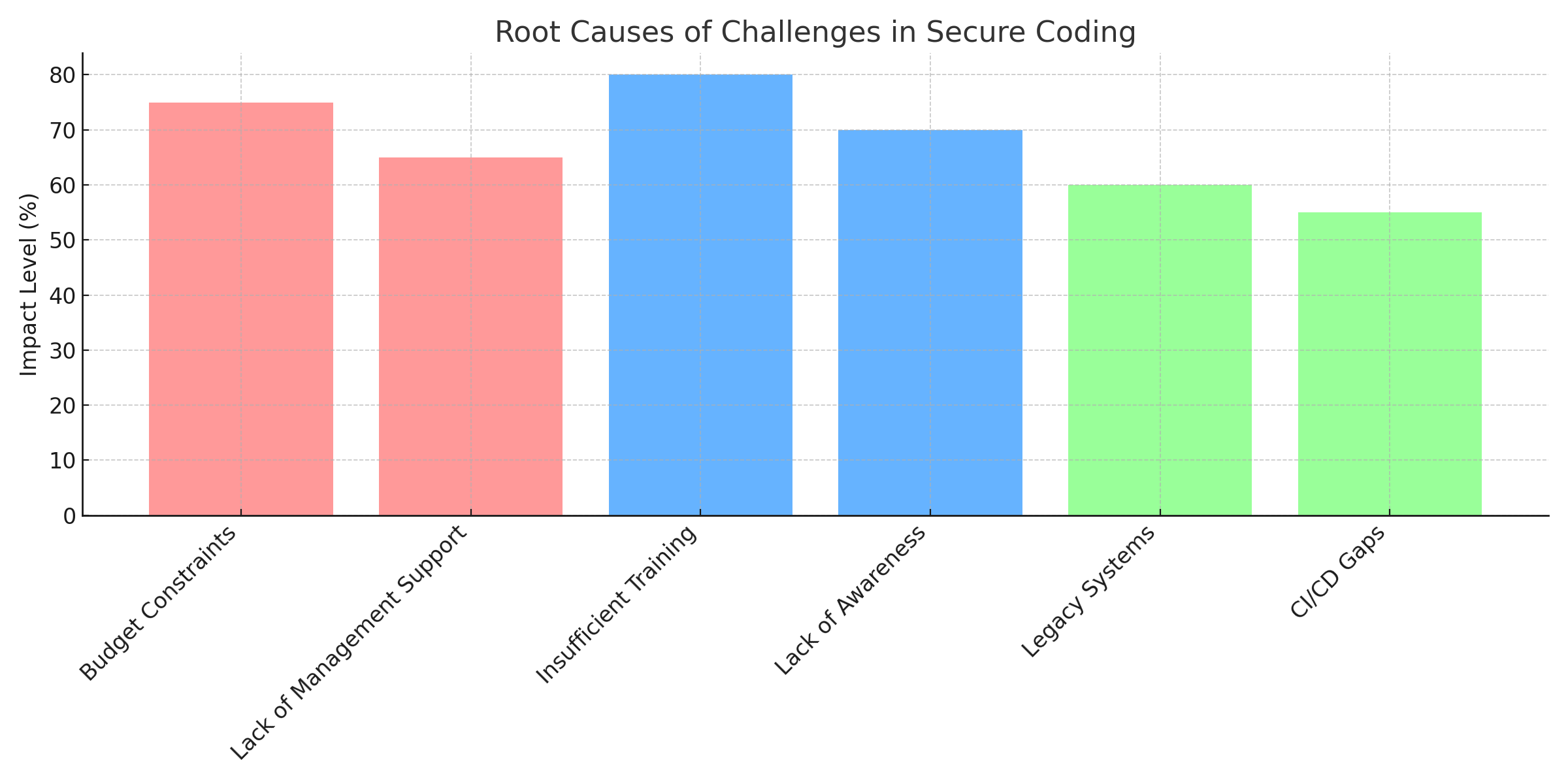}
    \caption{Root Causes of Challenges in Secure Coding, grouped by Organizational, Educational, and Technical categories.}
    \label{fig:root_causes}
\end{figure}

To better visualize the prevalence and grouping of implementation barriers, Figure~\ref{fig:root_causes} presents a summary of root causes across organizational, educational, and technical domains.

\section{Emerging Directions in Secure Coding}

As threats grow in sophistication and software stacks become more agile and distributed, several directions in secure coding have gained prominence to close the gap between legacy practices and modern delivery. This section highlights three influential movements: AI-enabled security automation, zero-trust-by-default access control, and codified security within DevSecOps pipelines.

\subsection{AI-Enabled Security Automation}

Artificial Intelligence (AI) increasingly augments software security by automating checks across the development lifecycle. AI-based tools can traverse codebases to surface vulnerabilities, flag anomalous behaviors, and even propose or apply patches with near–real-time feedback~\cite{b12}. Offloading routine verification to automation reduces manual load and scales both speed and breadth of detection.

Sutton et al.~\cite{b15} report notable gains in test efficiency and coverage especially for large, complex systems. At the same time, McGraw~\cite{b2} warns against uncritical reliance: automated systems can misclassify issues or miss context-dependent logic flaws, so human review remains essential to validate findings and avoid blind spots.

While promising, AI-driven defenses are still maturing. Effective use requires continuous evaluation, explicit human oversight for high-impact decisions, and regular model updates to track evolving adversary techniques.

\subsection{Zero-Trust-by-Default Access Control}

Zero trust assumes no implicit trust for users or services inside or outside the perimeter. Every request must be authenticated, authorized, and continuously verified before access is granted~\cite{b7}. This stance matches today’s cloud-native, distributed environments that routinely handle sensitive data.

Shostack~\cite{b17} notes that combining sound secure-coding practices with zero-trust building blocks e.g., role-based access control, network segmentation, and ongoing verification curtails lateral movement after compromise and hardens systems against tampering. The trade-off is operational complexity: short-lived credentials and strict policies can affect flow and performance~\cite{b33}, and success depends on skilled teams and well-integrated tooling.

\subsection{Codified Security in DevSecOps Pipelines}

Security-as-Code (SaC) captures policies, validation rules, and configurations directly in source and infrastructure-as-code artifacts. As a core DevSecOps practice, SaC embeds security gates throughout CI/CD so that every build and deploy enforces unified controls~\cite{b4,b5}.

By expressing controls as code, organizations apply consistent checks access policies, vulnerability scans, and compliance tests early and repeatedly. As shown in our prior work~\cite{b44}, remote-first workflows and their communication patterns are tightly coupled with adopting such integrated practices. SaC helps surface issues sooner and reduce remediation cost~\cite{b34,b1}, commonly including guarded deployments, RBAC definitions, and automated security testing frameworks.

\subsection{Synthesis of Trends}

Taken together, these movements shift security from reactive add-ons to proactive, embedded, and scalable mechanisms. AI automation accelerates detection, zero-trust reframes access control, and Security-as-Code operationalizes defenses inside CI/CD. Used in concert and with awareness of their limits and governance needs they provide a durable foundation for securing modern applications.

\section{Evaluating LLMs for Secure Code Review: A Case Study}

Large Language Models (LLMs) have found applications across diverse domains such as software development \cite{b41, b42}, healthcare \cite{b39, b40}, and education \cite{b43}. In software engineering, they are increasingly being explored for tasks like code generation ~\cite{b35, b36}, documentation, and review. In the context of secure coding, these models may also offer potential as lightweight vulnerability detectors during early development stages ~\cite{b37, b38}. To assess this capability, we conducted a case study to evaluate how effectively an LLM can identify common security issues in code.

\subsection*{A. Methodology}

We selected three intentionally vulnerable code snippets in commonly used web development languages (Python and JavaScript), each representing a typical flaw: SQL injection (SQLi), cross-site scripting (XSS), and broken authentication. The prompts were submitted to publicly accessible interface of GPT (ChatGPT with GPT-4), asking:

\textit{"Analyze this code and describe any security vulnerabilities."}

The LLM’s responses were analyzed for correctness, depth, and completeness.

\subsection*{B. Results}

Table~\ref{tab:llm-results} summarizes the results of the LLM based vulnerability analysis. As shown, the model was able to accurately identify and explain a classic SQL injection flaw in Python code. For the XSS example in JavaScript, the LLM correctly flagged the issue but provided only a partial explanation, lacking detail on encoding or proper output sanitization. In contrast, the model failed to recognize a broken authentication scenario, highlighting current limitations in detecting logic based vulnerabilities that require deeper contextual understanding.

\begin{table*}
\centering
\caption{LLM Detection of Code Vulnerabilities.}
\setlength{\tabcolsep}{5pt}
\renewcommand{\arraystretch}{1.1}
\begin{tabular}{|p{4.5cm}|p{2.5cm}|c|c|}
\hline
\textbf{Vulnerability Type} & \textbf{Language} & \textbf{LLM Detected?} & \textbf{Explanation Accurate?} \\
\hline
SQL Injection & Python & Yes & Yes \\
Cross-Site Scripting (XSS) & JavaScript & Yes & Partially \\
Broken Authentication & JavaScript & No & N/A \\
\hline
\end{tabular}
\label{tab:llm-results}
\end{table*}

\subsection*{C. Discussion}

The results show that LLMs can reliably detect certain syntactic vulnerabilities, such as unsanitized SQL queries and basic XSS vectors. However, more nuanced and contextual issues like broken authentication logic were not accurately identified. The LLM often provided generic suggestions without fully understanding the logic or flow of the code. These findings highlight both the strengths and current limitations of LLMs in the secure coding domain.

While promising as support tools for early review, LLMs should not be relied upon as primary security evaluators. They are best viewed as complementary aids that may assist developers, particularly those with limited security experience, in identifying obvious flaws before formal testing or peer review.

\subsection*{D. Comparison with Traditional Security Tools}

To contextualize the strengths and limitations of large language models in secure code analysis, Table~\ref{tab:llm-vs-tools} presents a structured comparison between LLMs and traditional static and dynamic analysis tools. The comparison covers setup complexity, detection capabilities, explainability, and integration potential in development workflows.

\begin{table}
\caption{Comparison of LLMs and Traditional Static/Dynamic Analysis Tools}
\centering
\resizebox{\columnwidth}{!}{%
\begin{tabular}{|p{3.3cm}|p{3.3cm}|p{3.3cm}|}
\hline
\textbf{Feature} & \textbf{LLMs (e.g., ChatGPT)} & \textbf{Traditional Tools} \\
\hline
Setup Time & None (zero setup) & Requires installation and configuration \\
\hline
Input Format & Natural language or code snippets & Source code, binaries, or compiled artifacts \\
\hline
Output Type & Human-readable explanation & Structured reports with warnings/errors \\
\hline
Context Understanding & Partial, limited to prompt scope & Strong syntactic analysis, limited semantic depth \\
\hline
Detection of Logic Flaws & Limited & Possible with dynamic tools, but not guaranteed \\
\hline
False Positive Rate & Low to Medium & Medium to High (especially in static tools) \\
\hline
CI/CD Integration & Manual or external scripting required & Built-in support in modern DevSecOps pipelines \\
\hline
Learning Curve & Low (conversational interface) & Medium to High, depending on tool complexity \\
\hline
Explainability of Results & High (natural language reasoning) & Often limited to error codes or technical traces \\
\hline
\end{tabular}
}
\label{tab:llm-vs-tools}
\end{table}

\section{Future Research Directions}

Despite notable progress in secure coding, the pace of software change, shifting threat landscapes, and the growth of distributed architectures keep revealing new weaknesses. To close these gaps, further work is needed to overcome limits in existing frameworks and make them fit modern environments. We outline three priorities: education, automation, and adaptation to emerging technologies.

\subsection*{A. Advancing Secure Coding Education}

A core but still underdeveloped need is strengthening secure coding education in universities and professional training. Many new engineers encounter security late, increasing the risk of early lifecycle vulnerabilities~\cite{b10,b3}. McGraw~\cite{b2} argues that security should be treated as a design tenet from the outset, not a downstream fix.

Future studies should define standardized curricula at the undergraduate and graduate levels, aligned with widely used frameworks such as OWASP and NIST. Empirical evaluations are also needed to measure the impact of hands-on approaches (e.g., cyber ranges and CTF-style exercises) on sustained security proficiency in both students and practitioners.

\subsection*{B. Next-Generation Security Automation}

Today’s automated tools help catch common issues but often suffer from false positives and limited understanding of contextual logic. Enhancing these systems with richer machine learning and natural language capabilities could improve precision, especially for semantic and business-logic flaws~\cite{b12,b15}.

As Sutton et al.~\cite{b15} note, fuzzing and automated scanning can uncover subtle bugs, yet expert interpretation remains essential. Future work should prioritize explainable AI within security tooling models that not only detect but justify findings and seamless CI/CD integration to provide near real-time checks during rapid releases.

\subsection*{C. Adapting Practices to Emerging Architectures}

The uptake of microservices, serverless, and cloud-native platforms introduces distinct security challenges. Traditional guidance can falter in highly dynamic, decentralized environments~\cite{b8}. Ephemeral services, infrastructure-as-code, and third-party dependencies demand more adaptive methods.

Building on Shostack~\cite{b17}, research should deepen cloud-tailored threat modeling and automation. Extending SSDLC to DevSecOps workflows using Security-as-Code to enforce continuous policy and compliance deserves focused study. Further, embedding secure coding at the infrastructure layer (e.g., within Kubernetes and related orchestration) remains a key frontier.

\subsection*{D. A Living Research Agenda}

Going forward, progress hinges on a balance of foundational education, trustworthy automation, and architectural adaptability. Addressing these dimensions will strengthen web application security while keeping pace with modern engineering practice. Collaboration across academia, industry, and open-source ecosystems will be crucial to deliver practical, scalable, and effective defenses for next-generation systems.

\section{Conclusion}

Secure coding remains a foundational discipline in protecting web applications from increasingly sophisticated cyber threats. This paper provided a comprehensive review of key secure coding frameworks including the OWASP Top 10, the NIST Cybersecurity Framework, and the Secure Software Development Lifecycle (SSDLC) all of which offer structured methodologies for reducing software vulnerabilities~\cite{b6,b7,b4}. Despite their widespread availability, consistent adoption across organizations continues to face obstacles such as inadequate training, limited resources, and the rapid pace of software delivery~\cite{b1,b3,b9}.

To address these challenges, emerging trends such as AI-driven vulnerability detection, zero-trust security models, and Security-as-Code practices within DevSecOps pipelines offer promising directions~\cite{b12,b17,b5}. However, these approaches must be further refined and operationalized to ensure scalability and effectiveness in production environments.

Future research should focus on embedding secure coding principles into formal education, enhancing the intelligence and interpretability of automated security tools, and evolving traditional models to align with modern cloud native and distributed systems~\cite{b10,b15,b8}. Bridging these gaps will be essential for developing software systems that are not only functional and performant but also inherently secure by design.

By synthesizing foundational frameworks, current implementation challenges, and cutting edge trends, this work contributes a holistic perspective to the secure coding landscape offering valuable guidance to researchers, educators, and practitioners seeking to advance the state of secure software development.

\begin{itemize}
\item Secure coding remains foundational, yet under-adopted.
\item LLMs provide helpful early feedback but cannot replace expert review.
\item Embedding secure practices in education, automation, and cloud-native workflows is critical for future resilience.
\end{itemize}

\section*{Acknowledgment}
This paper originated from a group work in the Secure Software Engineering course in the Bowling Green State University. We thank Humphrey Borketey, Prabhjot Kaur, and Kennedy Kporxah for their contributions to the original class submission.

\section*{Author Contributions}
Kiana Kiashemshaki: Conceptualization, Methodology, Investigation, Literature review, Formal analysis, Visualization (figures/tables), Writing original draft, Writing review \& editing, Project administration.

Mohammad Jalili Torkamani: Methodology (LLM evaluation design), Experiments and data curation (case-study prompts/results), Validation, Writing review \& editing, Resources.

Negin Mahmoudi: Framework comparison synthesis (OWASP/NIST/SSDLC), Validation, Writing review \& editing, Visualization (framework/timeline figures).

\section*{Funding}
No dedicated funding was received for this project.

\section*{Conflict of Interest}
The authors declare that they have no conflicts of interest.

\clearpage
\appendix
\section*{Appendix: LLM Prompts and Responses}

This appendix presents the full prompts and LLM-generated responses used in the case study described in Section VI. Each prompt was submitted to ChatGPT (GPT-4), and the model's output is shown below without modification.

\subsection*{Case 1: SQL Injection}

\textbf{Prompt}
\begin{lstlisting}[language=Python, caption=Prompt for SQL Injection]
def login():
    username = request.args.get("user")
    password = request.args.get("pass")
    query = "SELECT * FROM users WHERE name = '" + username + "' AND pwd = '" + password + "'"
    cursor.execute(query)
\end{lstlisting}

\textbf{LLM Response}
\begin{lstlisting}[language=Python, caption=LLM Response for SQL Injection]
This code is vulnerable to SQL Injection attacks.

SQL Injection is a code injection technique that attackers can use to insert malicious SQL statements into input fields for execution. This can allow the attacker to view data that they are not authorized to access, manipulate or delete data, or even execute administration operations on the database.

In this code, the username and password are directly concatenated into the SQL query without any form of sanitization or parameterization. This means that an attacker could provide a username or password that contains SQL code, which would then be executed by the database.

For example, an attacker could use the username `admin' --` . This would cause the SQL query to become `SELECT * FROM users WHERE name = 'admin' --' AND pwd = ''`, effectively commenting out the password check and allowing the attacker to login as the admin user without knowing the password.

To mitigate this vulnerability, you should use parameterized queries or prepared statements, which can ensure that user input is always treated as literal data and not executable code. Here is an example of how you could modify the code to use parameterized queries in Python with a SQLite database:

def login():
    username = request.args.get("user")
    password = request.args.get("pass")
    query = "SELECT * FROM users WHERE name = ? AND pwd = ?"
    cursor.execute(query, (username, password))

In this version of the code, the `?` characters are placeholders that the `execute` method replaces with the `username` and `password` variables in a safe manner that prevents SQL Injection.
\end{lstlisting}

\subsection*{Case 2: Cross-Site Scripting (XSS)}
\textbf{Prompt}
\begin{lstlisting}[language=HTML, caption=Prompt for XSS]
<!DOCTYPE html>
<html>
  <body>
    <form method="GET">
      <input name="msg" />
      <button type="submit">Submit</button>
    </form>
    <p>Your message: <script>document.write(location.search.split('=')[1])</script></p>
  </body>
</html>
\end{lstlisting}

\textbf{LLM Response}
\begin{lstlisting}[language=HTML, caption=LLM Response for XSS]
The code above is a simple HTML form that takes a user's input and displays it on the page. However, it has a significant security vulnerability related to Cross-Site Scripting (XSS).

The vulnerability lies in the line where it writes the user's input directly to the document without any sanitization or encoding:

<script>document.write(location.search.split('=')[1])</script>

This line takes the value from the URL parameter 'msg' and writes it directly to the page. If a user enters a string that includes JavaScript code, that code will be executed when the page loads. This is a classic example of a reflected XSS vulnerability.

For example, an attacker could craft a URL like http://example.com/?msg=<script>malicious_code_here</script>. If someone visits this URL, the malicious JavaScript code will be executed.

To fix this vulnerability, the application should properly encode or sanitize user inputs before displaying them on the page. This can be done using various methods depending on the programming language and framework used. For JavaScript, one could use the `textContent` property or `innerText` property instead of `document.write()`, as these properties do not interpret the input as HTML or JavaScript code.
\end{lstlisting}

\subsection*{Case 3: Broken Authentication}

\textbf{Prompt}
\begin{lstlisting}[language=Python, caption=Prompt for Broken Authentication]
def login():
    username = request.form['username']
    user = db.find_user(username)
    if user:
        session['logged_in'] = True
        return "Welcome!"
    else:
        return "User not found"
\end{lstlisting}

\textbf{LLM Response}
\begin{lstlisting}[language=Python, caption=LLM Response for Broken Authentication]
The code provided is a simple login function that checks if a user exists in the database and then logs them in if they do. However, there are several security vulnerabilities present:

1. No Password Verification: The function does not check for a password. It only checks if the username exists in the database. Anyone who knows a valid username can log in.

2. No Input Sanitization: The function does not sanitize the input from the user. This leaves the application vulnerable to SQL Injection attacks. An attacker could potentially input malicious SQL code as the username to manipulate the database.

3. No Encryption: The function does not encrypt the username before storing it in the session. This could potentially expose sensitive user information.

4. No Brute Force Protection: The function does not limit the number of login attempts. This makes it vulnerable to brute force attacks, where an attacker tries many different usernames until they find one that works.

5. No Error Handling: The function does not handle any errors that may occur during the login process. This could potentially expose sensitive information about the system or the database.

6. No Use of HTTPS: If this function is used in a web application that does not use HTTPS, the username could be intercepted by an attacker.

7. Session Fixation: The function does not regenerate the session ID after successful login. This makes it vulnerable to session fixation attacks, where an attacker sets a victim's session ID before they log in.

8. No Logging: The function does not log any events. This makes it difficult to detect and respond to security incidents.
\end{lstlisting}

\end{document}